\documentclass[twocolumn,showpacs,superscriptaddress,preprintnumbers,amsmath,amssymb.pdffig,floatfix,prb]{revtex4}
\usepackage{graphicx}
\usepackage{dcolumn}
\usepackage{color}
\usepackage{bm}
\begin{document}
\title{Magnetism in small bimetallic Mn-Co clusters}	
\author{Shreemoyee Ganguly}
\affiliation{Department of Material Sciences,  S.N. Bose National Center for Basic Sciences,  JD Block,  Sector III,  Salt Lake City,  Kolkata 700 098,  India } 
\author{Mukul Kabir}
\altaffiliation{Author to whom correspondence should be addressed}
\affiliation{Department of Material Sciences,  S.N. Bose National Center for Basic Sciences,  JD Block,  Sector III,  Salt Lake City,  Kolkata 700 098,  India }
\affiliation{Department of Materials Science and Engineering, Massachusetts Institute of Technology, Cambridge, Massachusetts 02139, USA }
\author{Soumendu Datta}
\affiliation{Department of Material Sciences,  S.N. Bose National Center for Basic Sciences,  JD Block,  Sector III,  Salt Lake City,  Kolkata 700 098,  India }
\author{Biplab Sanyal}
\affiliation{Division of Materials Theory, Department of Physics and Materials Science, Uppsala University, Box 530, SE-75121 Uppsala, Sweden.}
\author{Abhijit Mookerjee}
\affiliation{Department of Material Sciences,  S.N. Bose National Center for Basic Sciences,  JD Block,  Sector III,  Salt Lake City,  Kolkata 700 098,  India }

\date{\today}
\begin{abstract}
Effects of alloying on the electronic and magnetic properties of Mn$_{x}$Co$_{y}$ ($x+y$=$n$=2-5; $x$=0-$n$) and Mn$_2$Co$_{11}$ nanoalloy clusters are investigated using the density functional theory (DFT).   Unlike the bulk alloy, the Co-rich clusters are found to be ferromagnetic and the magnetic moment increases with Mn-concentration, and is larger than the moment of pure Co$_n$ clusters of same size.  For a particular sized cluster the magnetic moment increases by 2 $\mu_B$/Mn-substitution, which is found to be independent of the size and composition. All these results are in good agreement with recent Stern-Gerlach (SG) experiments [Phys. Rev. B {\bf 75}, 014401 (2007) and Phys. Rev. Lett. {\bf 98}, 113401 (2007)]. Likewise in bulk Mn$_x$Co$_{1-x}$ alloy, the local Co-moment decreases with increasing Mn-concentration. 
\end{abstract}

\maketitle

\section{Introduction}

Nanoalloy clusters have received considerable attention for their peculiar catalytic, optical, magnetic, electronic,  and geometric  properties. \cite{Jellinek1999Springer, Kondow2003, Klabunde1994,  Andrews1992, Jellinek2008Faraday} For such clusters, chemical and physical properties can be tailored by varying not only the size but also the composition for a specific purpose. This opens the way to a large variety of potential applications in areas such as high-density recording,\cite{Sun200} catalysis,\cite{Sinfelt1983, Molenbroek2001, Molenbroek1997} optics,\cite{Shibata2002, Darby2002, Ruban1999} and biomedical.\cite{Loo2005} Particularly, the interest in transition metal clusters arises from a desire to seek a solution to the technologically important question:  how magnetic properties change in reduced dimensions? The candidates chosen for the present study, Mn and Co have very interesting properties in low dimensions. Manganese, though antiferromagnetic as bulk, shows finite magnetic moment in reduced dimension\cite{Mark2001, Mark2004, Kabir2006, Kabir2007} whereas, Cobalt shows enhanced magnetic moment compared to the bulk.\cite{Xu2005,Mark2006,Datta2007} Therefore, it will be interesting to see how the properties of the bimetallic cluster formed out of these two elements change with composition, atomic ordering, and size. 

The first step to the study of cluster properties is the determination of the ground state structures and the complexity to locate that increases with the cluster size as the number of local minima in the potential energy surface increases. This leaves a number of possible geometric and/or magnetic isomers in pure clusters for each size. Compared to the pure clusters, in alloy clusters ``homotops"\cite{Jellinek1999Springer} are possible in addition to such usual isomers. Jellinek introduced the term ``homotop" to describe A$_x$B$_y$ alloy cluster isomers, with a fixed size ($x+y$=$n$) and composition ($x/y$ ratio), which have the same geometrical arrangement of atoms, but differ in the way A- and B-type atoms are arranged. Due to the presence of these homotops, there arises a large number of combinatorial possibilities which makes the finding of lowest energy structures for alloy clusters even more computationally expensive task than that for pure clusters. Thus most of the theoretical studies done on bimetallic alloy clusters take resort to some empirical manybody potential to reduce the computational expense.\cite{Jellinek1999Springer, Christensen1995, Lopez1996, Jellinek1996, Krissinel1997CPL, Krissinel1997IJQC,Jellinek1999} Our study is one of the  very few which uses  an {\it ab-initio} methodology for transition metal nanoalloy clusters. Here we study bimetallic Mn$_{x}$Co$_{y}$ clusters in the size  ranging  from 2 to 5 for all possible stoichiometry. Using first-principles DFT we study  the evolution of structural, electronic and magnetic properties as we change the size and composition. The interplay between these properties yield many interesting features. Such features are analyzed in greater depth through density of states and partial charge density. 

In earlier works, we have studied pure Mn$_n$ (Refs.\cite{Kabir2006, Kabir2007}) and Co$_n$ (Ref. \cite{Datta2007}) clusters. A transition from ferromagnetic to ferrimagnetic Mn-Mn coupling is observed at $n=$ 5 for Mn$_n$ clusters and the ferrimagnetic states continue to be the ground state for larger clusters. On the other hand, pure Co$_n$ clusters are found be ferromagnetic with magnetic moment higher than the bulk. Calculated magnetic moments of pure Mn$_n$ and Co$_n$ clusters show very good agreement with the Stern-Gerlach  molecular beam experiments.\cite{Mark2001,Mark2004,Xu2005,Mark2006}
 
Neutron scattering studies of bulk Mn$_{x}$Co$_{1-x}$ alloy have been used to determine the variation of individual atomic moments, $\mu_{\rm Mn}$ and  $\mu_{\rm Co}$, with increasing Mn:Co ratio.\cite{Mark2007, Cable1982, Menshikov1985} On the Co-rich side,  $\mu_{\rm Mn}$ and  $\mu_{\rm Co}$ are aligned antiferromagnetically, with the magnitude of both $\mu_{\rm Co}$ and $\mu_{\rm Mn}$ decreasing monotonically with increasing Mn content such that the mean per-atom  moment ($\bar{\mu}$) of the alloy also decreases strongly with Mn content. Infact, $\bar{\mu}$ decreases from 1.72 $\mu_B$ for $x$=0 to 0 $\mu_B$ for $x$=0.32 (Ref. \cite{Matsui1970}). However, small Mn-Co clusters have been found to behave in a completely different way.\cite{Mark2007,Yin2007} Recently, Knickelbein has found that unlike bulk Mn$_{x}$Co$_{1-x}$ alloys, in which the presence of Mn forces the mean per-atom moment to decrease, the significant presence of Mn in medium sized Mn$_x$Co$_y$ ($n$=$x+y$=11$-$29) clusters results in overall magnetic moment that are comparable to those of the corresponding pure Co$_n$ clusters, and in some cases (e.g. $n$=11-14) even larger. More recently Yin {\it et al.} have measured the magnetic moments of Mn$_x$Co$_y$ ($y\le$60; $x$$\le$$y/3 $) clusters and found an increase in per-atom magnetic moment for Co-rich Mn$_x$Co$_y$ cluster with increasing Mn concentration.\cite{Yin2007} This enhancement in moment due to Mn doping is independent of cluster size and composition. On the other hand, for Mn-rich clusters, for more than 40\% Mn concentration, the average magnetic moment of  Mn$_x$Co$_y$ cluster decreases with increasing Mn concentration. This suggests that unlike bulk Mn-Co alloys, both Mn and Co within the small Mn-Co clusters retain substantial moments even at high Mn fractions. However, the magnitudes of individual Mn and Co moments could not be measured, and consequently the nature of magnetic coupling can not be concluded in a SG experiment. Motivated by these recent SG experiments,\cite{Mark2007, Yin2007} in the present paper, we study Mn$_{x}$Co$_{y}$ clusters from first-principles in the size range $n$=2-5. In order to make a direct comparison to the experiments  we also study Mn$_{2}$Co$_{11}$ cluster, which lies within the experimental regime.\cite{Mark2007}

\section{Computational Details}               

Calculations are performed using DFT based  pseudopotential plane wave method.\cite{Kresse1999} We have chosen the  projector augmented wave  method, \cite{Blochl1994} and used the Perdew-Bruke-Ernzerhof exchange-correlation functional\cite{Perdew1996} for the spin-polarized generalized gradient correction. The 3$d$ and 4$s$ electrons of Mn and Co are treated as valence electrons. The wave functions are expanded in  a plane wave basis set within  270 eV kinetic energy. Reciprocal space integrations are carried out at the ${\Gamma}$ point. Symmetry unrestricted optimizations (of both geometry and spin) are performed using the conjugate gradient and quasi-Newtonian methods until all the force components are less than a threshold value of 0.005 eV/\r{A}. Simple cubic supercells are used with periodic boundary conditions, and it is made sure that two neighboring clusters are separated by at least 10 \r{A} vacuum space. This ensures that the interaction of a cluster with its periodic image is negligible. Earlier we have used same methodology to study pure Mn$_n$ and Co$_n$ clusters.\cite{Kabir2006, Datta2007} For each cluster size all possible ``homotops'' have been considered for several geometric structures with all possible 
compositions. We have also considered all possible spin multiplicities for each of these structures. These ensures the robustness for the ground state search. It should be mentioned here that calculations are done within the collinear spin assumption.

The binding energy per atom ($E_{B}$) is defined as,

\begin{equation}
E_B({\rm Mn}_{x}{\rm Co}_{y})=\frac{1}{n}[xE({\rm Mn})+yE({\rm Co})-E({\rm Mn}_{x}{\rm Co}_{y})],
\end{equation}

where $x$ ($y$) is the number of Mn (Co) -atoms  in Mn$_x$Co$_y$  cluster, $n$(= $x+y$) is the cluster size  and  $E({\rm Mn}_{x}{\rm Co}_{y})$, $E({\rm Mn}$), and  $E({\rm Co})$ are the total energies of Mn$_{x}$Co$_{y}$ cluster, and an isolated Mn- and Co-atom, respectively. For a given $n$ and for a certain composition, the structure with  the highest binding energy is considered to be the {\it ground state}. 
The local magnetic moment $\mu_{\rm X}$ at X-atom can be calculated from,
\begin{equation}
\mu_{\rm X}=\int_0^R[\rho_{\uparrow}({\mathbf r})-\rho_{\downarrow}({\mathbf r})]\,d\mathbf{r}
\end{equation}
where $\rho_{\uparrow}({\mathbf r})$ and $\rho_{\downarrow}({\mathbf r})$ are spin-up and spin-down charge-densities, respectively and $R$ is the radius of the sphere centered on the atom X. For a particular cluster, $R$ is taken such that no two spheres overlap i.e. $R$ is equal to half of the shortest bond length in that cluster.

\section{Results and Discussions}

\subsection{\label{gsandisomer} Ground states and significant isomers}

It is necessary to carry out calculations for not only the ground state, but also for the low energy isomers, i.e. clusters with different geometries, homotops and different magnetic arrangements, which have energies close to that of the ground state. This is because when, in our earlier works,\cite{Kabir2006, Datta2007} theoretical results were compared with experimental results, it was noted that for a particular size of cluster, the isomers with different magnetic moments are likely to be present in the  SG beam with a statistical weight and essentially the measured magnetic moment is the weighted average of the moments of all those isomers.  Previously we have extensively studied the pure Mn$_n$ (Ref. \cite{Kabir2006}) and Co$_n$ (Ref. \cite{Datta2007}) clusters within the same theoretical methodology. Therefore, we do not elaborate pure clusters here, rather we refer the  readers to Ref.\cite{Kabir2006} and Ref.\cite{Datta2007}.

{\it Mn$_x$Co$_y$ ($x+y$ = 2):} Due to the half-filled 3$d$ and filled 4$s$ states  and due to high 4$s^2$3$d^5$ $\rightarrow$ 4$s^1$3$d^6$ promotion energy Mn-atoms bind very weakly when they are brought together. The binding energy of Mn$_2$ dimer is 0.52 eV/atom and the bond length is comparatively higher than all other 3$d$ transition metal dimers.\cite{Kabir2006} On the other hand, Co$_2$ dimer has much higher binding energy (1.45 eV/atom) and smaller bond length (1.96 \AA).\cite{Datta2007} Both the pure dimers are ferromagnetic with 10 (Mn$_2$) and 4 $\mu_B$ (Co$_2$) moments. The bond length of MnCo dimer is in between the pure dimers and the binding energy increases monotonically in Mn$_2$$<$MnCo$<$Co$_2$ order (Table \ref{table:n=2}). The Mn-Co coupling is also ferromagnetic in the mixed dimer. 
   
\begin{table}[t!]
\caption{\label{table:n=2}Structure, binding energy $E_B$, relative energy to the ground state (GS), $\Delta$$E$=$E-$$E_{{\rm GS}}$, per-atom magnetic moment   $\bar{\mu}$ and average bondlength $\langle L_B \rangle$ for the ground states of Mn$_x$Co$_y$ clusters with $x+y = 2$.}
\begin{tabular}{lccccc}
\hline
\hline
Cluster &  Structure & $E_B$ & $\Delta E$ & $\bar{\mu}$  & $ \langle L_B \rangle$ \\
        &            & (eV/atom) & (eV) & ($\mu_B$/atom) & (\AA) \\ 
\hline
\hline
Co$_2$ & linear & 1.45 & 0.00 & 2 & 1.96 \\
MnCo   & linear & 1.09 & 0.00 & 3 & 2.06 \\
Mn$_2$ & linear & 0.52 & 0.00 & 5 & 2.58 \\
\hline
\end{tabular}
\end{table}

\begin{figure}[!b]
\includegraphics[width=8cm,  keepaspectratio]{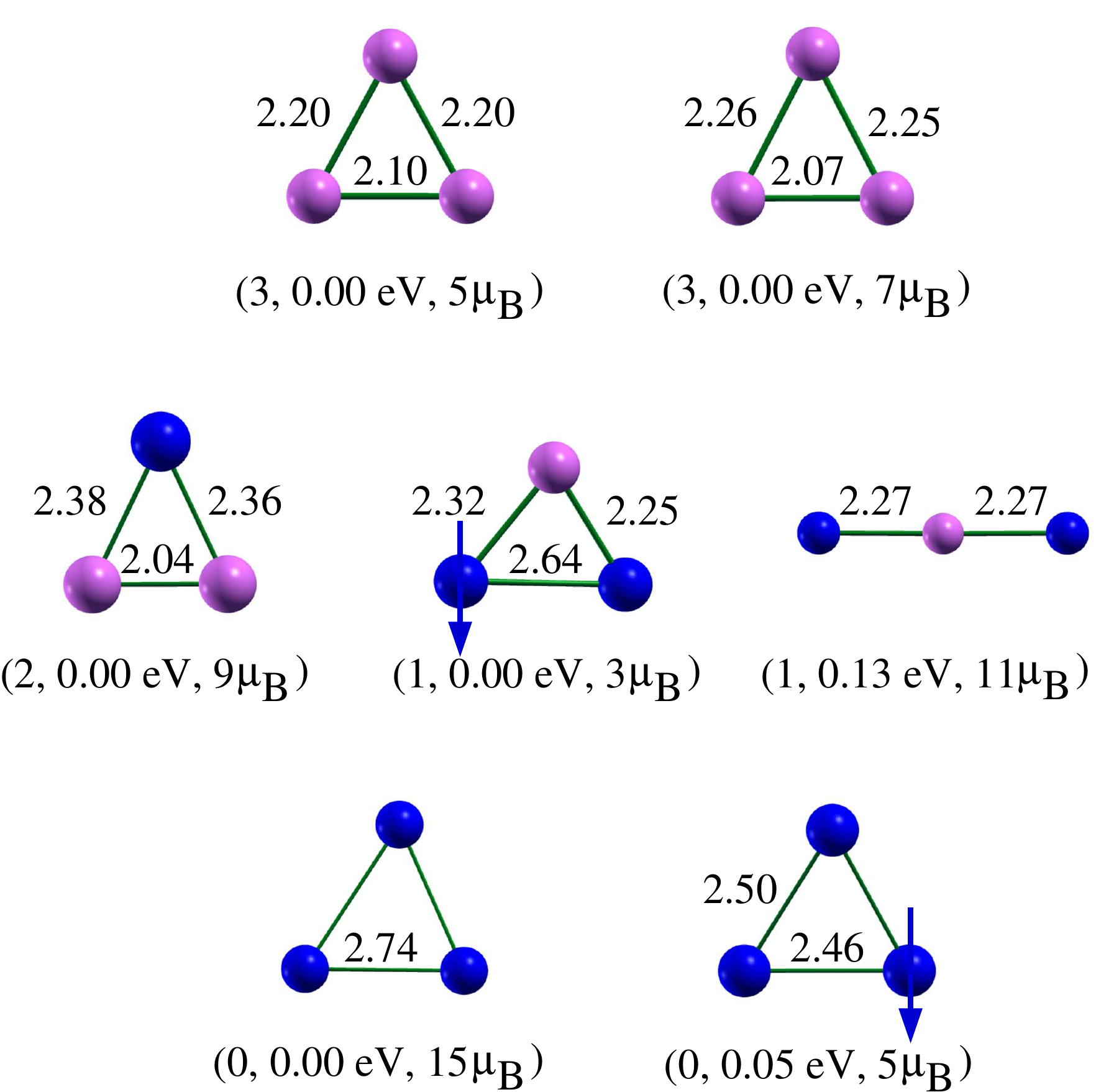}
\caption{\label{fig:trimer}(Color online) Ground state structures and significant isomers of Mn$_x$Co$_y$ ($x+y=3$) clusters. Lighter (Magenta) shades represent Co-atoms and darker (blue) shades represent Mn-atoms. The bond lengths are quoted in \AA. The numbers in the parenthesis represent the number of Co-atoms ($y$),  difference in total energy from the corresponding ground state ($\Delta E$), and total magnetic moment, respectively. Down arrow represents the atom with antiparallel moment.}
\end{figure}

{\it Mn$_x$Co$_y$ ($x+y=3$):}  
The linear and triangular structures were taken as initial guess and the results are shown in Fig.\ref{fig:trimer} and Table \ref{table:n=3}. The ground state structure are found to be  triangular for all the compositions. The pure Co$_3$ has two degenerate ground states. The magnetic moment is enhanced by 2 and 4 $\mu_{B}$ from the two degenerate ground state structures of Co$_3$ due to single Mn doping.  For Mn$_2$Co  the lowest energy state with a stable spin distribution\cite{stable} is ferrimagnetic,  with one of the Mn atoms having opposite spin to the other Mn and Co. The closest isomer to this is a linear ferromagnetic structure.   However, it lies 0.39 eV above the ground state. Here the Mn-Mn distance is large (4.54 \AA) causing them to be ferromagnetically coupled. The  ferromagnetic ground  state of pure Mn$_3$ was found to be nearly degenerate with a ferrimagnetic structure.\cite{Kabir2006}

\begin{table}[!t]
\caption{\label{table:n=3} Same as Table \ref{table:n=2} for $x+y=3$ clusters.}
\begin{tabular}{lccccc}
\hline
\hline
Cluster &  Structure & $E_B$ & $\Delta E$ & $\bar{\mu}$ & $ \langle L_B \rangle$ \\
        &            & (eV/atom) & (eV) & ($\mu_B$/atom) & (\AA) \\ 
\hline
\hline
Co$_3$ & Triangular & 1.78 & 0.00 & 1.67 & 2.16 \\
 & Triangular & 1.78 & 0.00 & 2.33 & 2.19\\
MnCo$_2$   & Triangular & 1.67 & 0.00 & 3.00 & 2.26 \\
Mn$_2$Co & Triangular & 1.16 & 0.00 & 1.00 & 2.40 \\
             & Linear   & 1.12 & 0.13 & 3.67 & 2.27\\
     & (with endon Mn)&     &       &    &   \\
Mn$_3$ & Triangular & 0.82 & 0.00 & 5.00      & 2.74  \\
 & Triangular & 0.80 & 0.05 & 1.67 & 2.48 \\  
\hline
\end{tabular}
\end{table}

\begin{table}[!b]
\caption{\label{table:n=4}Same as Table \ref{table:n=2} for $x+y=4$ clusters.}
\begin{tabular}{lccccc}
\hline
\hline
Cluster &  Structure & $E_B$ & $\Delta E$ & $\bar{\mu}$ & $ \langle L_B \rangle$ \\
        &            & (eV/atom) & (eV) & ($\mu_B$/atom) & (\AA) \\ 
\hline
\hline
Co$_4$ & Tetrahedral & 2.28 & 0.00 & 2.5 & 2.34 \\
  & Rhombus & 2.25 & 0.11 & 2.5 & 2.25 \\
MnCo$_3$& Tetrahedral & 2.06 & 0.00 & 3.0 & 2.34 \\
 & Quadrilateral & 2.05 & 0.05 & 3.0 & 2.32\\
 & Tetrahedral & 2.02 & 0.18 & 2.5 & 2.31\\
 & Tetrahedral & 2.00 & 0.25 & 3.5 & 2.35\\ 
Mn$_2$Co$_2$ & Tetrahedral & 1.87 & 0.00 & 3.5 & 2.41 \\
 & Tetrahedral & 1.82 & 0.22 & 4.0 & 2.46\\
Mn$_3$Co & Tetrahedral & 1.60 & 0.00 & 4.0 & 2.54\\
 & Tetrahedral & 1.54 & 0.22 & 4.5 & 2.60\\
& Rhombus & 1.51 & 0.35 & 1.5 & 2.39\\
Mn$_4$ & Tetrahedral & 1.18 & 0.00 & 5.0 & 2.54\\
 & Tetrahedral & 1.16 & 0.08 & 2.5 & 2.59\\
 & Tetrahedral & 1.13 & 0.20 & 2.0 & 2.58\\
 & Tetrahedral & 1.13 & 0.20 & 0.0 & 2.65\\ 
\hline
\end{tabular}
\end{table}

\begin{figure}[!t]
\includegraphics[width=8cm,  keepaspectratio]{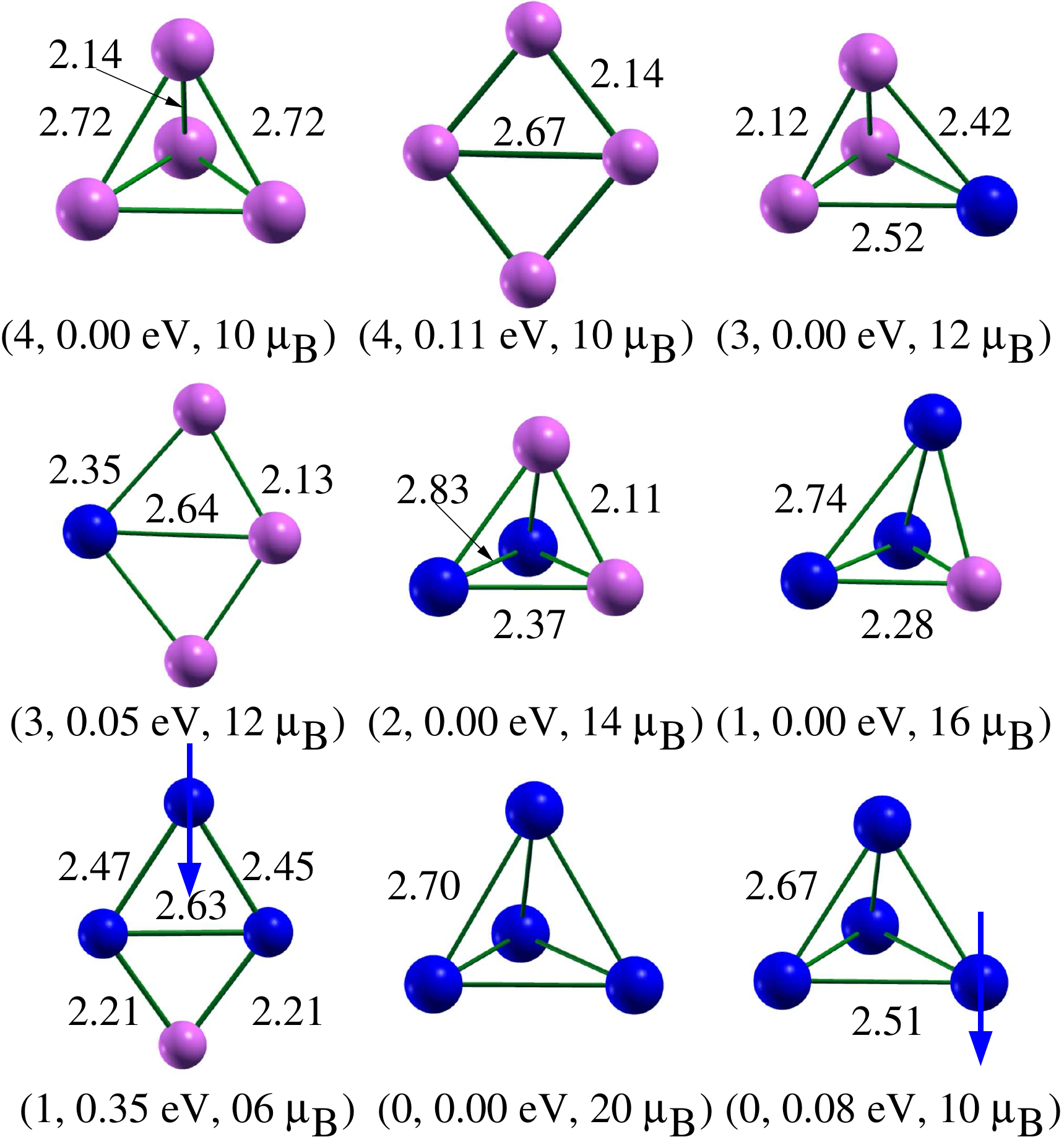}
\caption{\label{fig:tetramer}(Color online) Ground state structures and significant isomers of Mn$_x$Co$_y$ ($x+y=4$) nanoalloy clusters. Same conventions are followed as in Fig.\ref{fig:trimer}.}
\end{figure}
 
{\it Mn$_x$Co$_y$ ($x+y=4$):}  
For these clusters square and tetrahedral structres were considered. Results are summarized in Table \ref{table:n=4} and some of the relaxed structures are shown in Fig.\ref{fig:tetramer}. The ground state structures come out to be tetrahedral for all compositions. Co$_4$ has  a  Jahn-Teller distorted tetrahedral ground state\cite{Datta2007}  similar  to that  obtained by Castro {\it et al.},\cite{Castro1997} and the closest isomer is a  rhombus. Single Mn-substitution  (MnCo$_3$) increases the ground state moment by 2 $\mu_{B}$.  The first isomer is a planar structure,  a quadrilateral, having equal magnetic moment as the tetrahedral ground state. The next two closest isomers also have tetrahedral structure: one with average magnetic moment 0.5 $\mu_B$/atom lower and average bond length 0.03 \AA \ shorter and the other with average bond length 0.01 \AA \ longer and average magnetic moment 0.5 $\mu_B$/atom higher than that of the ground state. Thus the isomer with higher bond length has higher moment. This trend is seen for all compositions and sizes  (Tables I to IV). The closest non-ferromagnetic MnCo$_3$ isomer lies much higher (0.82 eV) than the ground state and has zero net moment. 

Substitution of another Co with Mn (Mn$_2$Co$_2$) further enhances the total magnetic moment  by 2 $\mu_{B}$ compared to MnCo$_3$ cluster.  In this case, the closest ferrimagnetic isomer, with average magnetic moment 1 $\mu_{B}$/atom,  lies far above (0.56 eV) from the ground state. A further substitution of a Co-atom with Mn atom (Mn$_3$Co) again leads to further 2 $\mu_{B}$ increase in magnetic moment.  A planar (rhombus)   ferrimagnetic isomer  lies 0.35 eV above the ground state. Mn$_4$ has binding energy 1.07 eV/atom, which is much lower than that of Co$_4$ and is ferromagnetic. However, it has a close ferrimagnetic isomer, which lies only 0.08 eV higher  in energy.  It is interesting to note here that for these clusters with $n$=4, the energy splitting between the ferromagnetic ground state and optimal ferrimagnetic state decreases with increasing Mn.

\begin{figure*}[t!]
\includegraphics[width=17cm,  keepaspectratio]{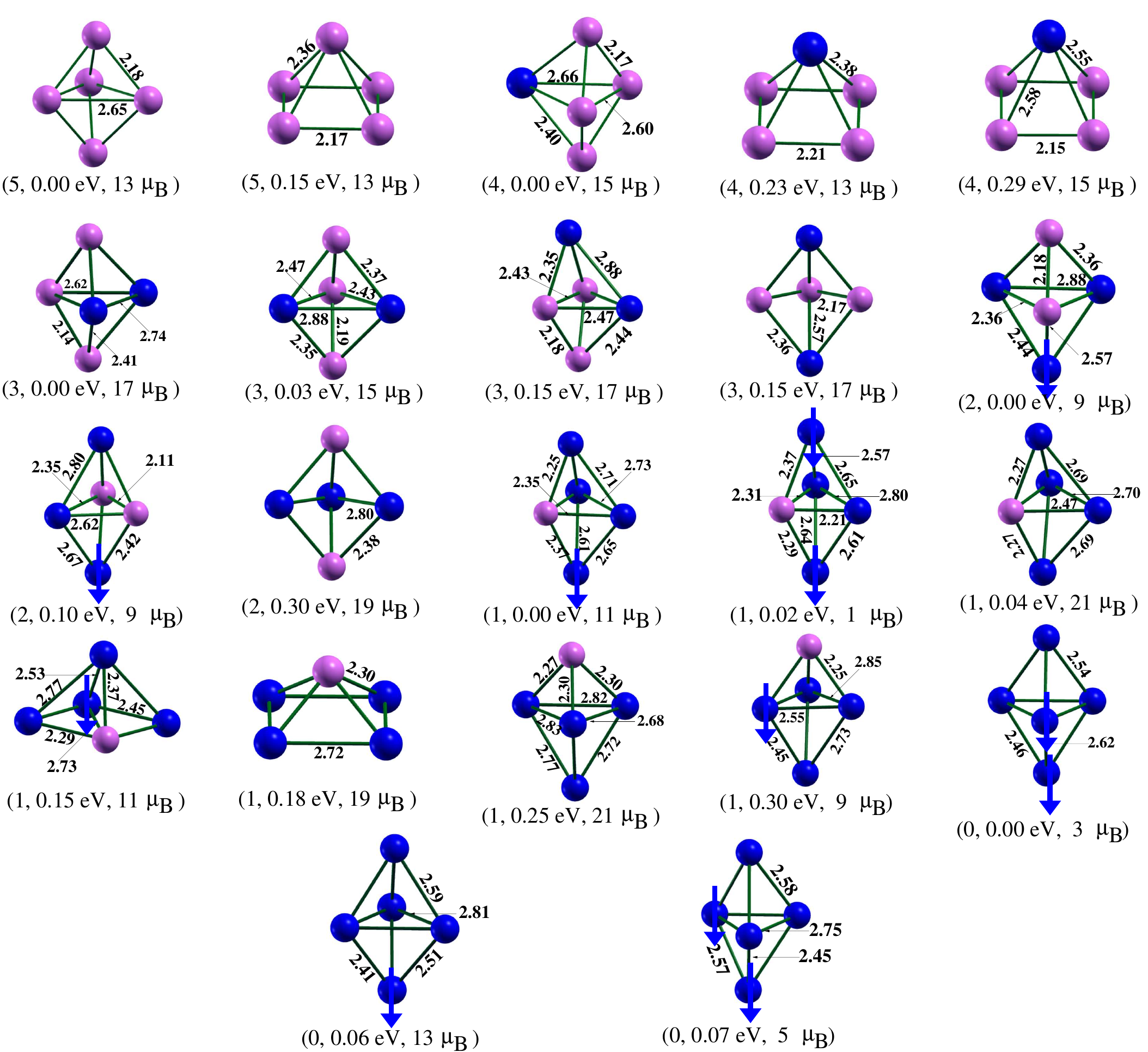}
\caption{\label{fig:pentamer}(color online) Ground state structures and significant isomers of Mn$_x$Co$_y$ ($x+y=5$) nanoalloy clusters. Same conventions are followed as in Fig.\ref{fig:trimer}.}
\end{figure*}

{\it Mn$_x$Co$_y$ ($x+y=5$):} 
The  trigonal bi-pyramidal (TBP), square pyramidal (SQPD) and (planar) pentagonal  geometries are considered as starting configurations. The results are tabulated in Table \ref{table:n=5} and minimum energy structures are shown in Fig.\ref{fig:pentamer}. All the clusters with different compositions have trigonal bi-pyramidal ground state. For MnCo$_4$ cluster, the ground state magnetic moment is enhanced by 2 $\mu_B$ compared to the pure Co-pentamer due to single Mn substitution.  The single-Mn doping also enhances the energy splitting between TBP and SQPD structures. We have found a similar trend in magnetic moment for Mn$_2$Co$_3$ as it is further enhanced by  2 $\mu_B$ due to another Mn-substitution. For this cluster the nearest `homotop' lies 0.15 eV higher. 

\vskip 0.2cm
\begin{table}[t]
\caption{\label{table:n=5}Same as Table \ref{table:n=2} for $x+y=5$ clusters.}
\begin{tabular}{lccccc}
\hline
\hline
Cluster &  Structure & $E_B$ & $\Delta E$ & $\bar{\mu}$ & $ \langle L_B \rangle$ \\
        &            & (eV/atom) & (eV) & ($\mu_B$/atom) & (\AA) \\ 
\hline
\hline
Co$_5$ & TBP & 2.55 & 0.00 & 2.6 & 2.34 \\
 & SQPD & 2.52 & 0.15 & 2.6 & 2.27\\
  
MnCo$_4$   & TBP & 2.46 & 0.00 & 3 & 2.38 \\
 & SQPD& 2.41 & 0.23 & 2.6 & 2.29\\
 & SQPD & 2.40 & 0.29 & 3 & 2.35\\

Mn$_2$Co$_3$& TBP & 2.25 & 0.00 & 3.4 & 2.43\\
 & TBP & 2.24 & 0.03 & 3 & 2.40\\
& TBP & 2.22 & 0.15 & 3.4 & 2.42\\
 & TBP & 2.22 & 0.15 & 3.4 & 2.37\\

Mn$_3$Co$_2$ & TBP & 2.03 & 0.00 & 1.8 & 2.45 \\
 & TBP & 2.01 & 0.10 & 1.8 & 2.45\\
 & TBP & 1.99 & 0.19 & 3.8 & 2.50\\
 & TBP & 1.99 & 0.22 & 3.4 & 2.44\\
 & TBP & 1.97 & 0.30 & 3.8 & 2.54\\

Mn$_4$Co & TBP & 1.76 & 0.00 & 2.2 & 2.52\\
 & TBP & 1.76 & 0.02 & 0.2 & 2.49\\
 & TBP & 1.75 & 0.04 & 4.2 & 2.55\\
 & TGPD & 1.73 & 0.15 & 2.2 & 2.52\\
 & SQPD & 1.72 & 0.18 & 3.8 & 2.51\\
 & TBP & 1.71 & 0.25 & 4.2 & 2.60\\
 & TBP & 1.70 & 0.30 & 1.8 & 2.51\\
 
Mn$_5$ & TBP & 1.41 & 0.00 & 0.6 & 2.54\\
 & TBP & 1.40 & 0.06 & 2.6 & 2.58\\
 & TBP & 1.40 & 0.07 & 1.0 & 2.59\\ 
\hline
\end{tabular}
\end{table}

The incremental behavior of magnetic moment with Mn-substitution is no longer observed as the clusters become ferrimagnetic, when the number of Mn-atoms increase to three or more, i.e., when the clusters become Mn-rich. Thus, magnetic moment of Mn$_3$Co$_2$ drops by 4 $\mu_B$ from the pure Co$_5$, as in this cluster one of the Mn-atoms  has antiparallel spin alignment with the others. The first isomer found is a homotop  of the ground state. A ferromagnetic homotop, which follows an increment of  2 $\mu_B$/Mn-substitution, lies much higher (0.30 eV), and has larger ($\sim$ 2.80 \AA)  Mn$_{\uparrow}$-Mn$_{\uparrow}$ separations.   In contrast, in the ferrimagnetic ground state Mn$_{\downarrow}$-Mn$_{\uparrow}$ distance is much shorter (2.44 \AA) than the Mn$_{\uparrow}$-Mn$_{\uparrow}$ distance (2.88 \AA ). This is in general true for other clusters (Fig.\ref{fig:pentamer}). The Mn$_4$Co is also ferrimagnetic, which has 2$\mu_{B}$ less moment than that of pure Co$_5$.  However, it has a ferromagnetic structure which obeys the  `2 $\mu_B$/Mn-substitution increment' rule and this structure lies only 0.04 eV higher. This cluster also has two different homotops in the form of a distorted tetragonal  pyramid (TGPD) and a SQPD and they lie 0.15 and 0.18 eV higher, respectively. It has previously been reported  that pure Mn$_5$ is ferrimagnetic in its ground state and a  ferromagnetic isomer lies 0.19 eV higher.\cite{Kabir2006}

\subsection{Binding energy and stability}

The coodination number increases with cluster size, and thus the binding  energy, which we have discussed earlier in detail for pure Mn$_n$ and Co$_n$ clusters.  \cite{Kabir2006, Kabir2007, Datta2007} However, the binding energy of pure Co$_n$ clusters are  much larger than that of pure Mn$_n$ clusters of same size. This is because the  Mn-atoms have half-filled 3$d$ and filled 4$s$ shells and also have high  $4s^{2}3d^{5}$ $\rightarrow$ $4s^{1}3d^{6}$ promotion energy. As a consequence,  the Mn-atoms do not bind strongly when they are brought together.\cite{Kabir2006, Kabir2007} 

\begin{figure}[!b]
\includegraphics[width=5.3cm, keepaspectratio, angle=0]{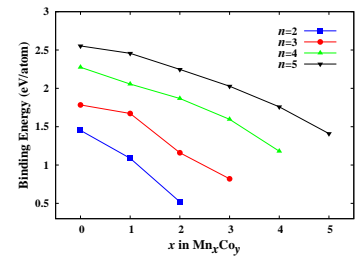}
\caption{\label{fig:ben}(color online) Binding energy of Mn$_x$Co$_y$ alloy clusters as a function of Mn-atoms.}
\end{figure}

\begin{figure}[!t]
\includegraphics[width=8.5cm, keepaspectratio, angle=0]{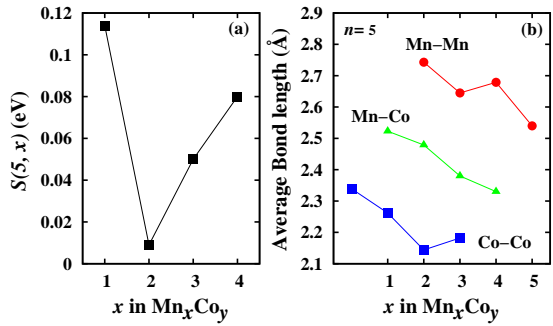}
\caption{\label{fig:sta}(color online) (a) Relative stability and (b) different (Mn-Mn, Mn-Co, and Co-Co) bond lengths of $n=5$ cluster as a function of Mn-atoms.}
\end{figure}

The Fig.\ref{fig:ben} shows that, in general, the binding energy decreases with increasing number of Mn-atoms ($x$) for a particular sized cluster $n$. This again is due to the weaker binding  among Mn atoms, and relatively weak Mn-Co binding than Co-Co bonding. 

The stability $S$ of these nanoalloy clusters with the variation of Mn-atoms $x$ in it can be defined as,
\begin{equation}
S(n,x) = E(n, x+1) + E(n, x-1) - 2E(n, x), 
\end{equation}
where $E(n,x)$ is the bound state energy of the Mn$_x$Co$_y$ ($n=x+y$) cluster, and is shown in Fig.\ref{fig:sta} for $n=5$. The MnCo$_4$ clusters is seen to have maximum  stability. A sudden dip in stability is seen for Mn$_2$Co$_3$, which can be described in terms of different bond distributions. We see that Mn$_2$Co$_3$ has maximum (minimum)  Mn-Mn (Co-Co) bond length,  and thus, has the weakest (strongest)  Mn-Mn (Co-Co) interaction among all the $n$=5 clusters with different compositions. This reduced hybridization of the Mn-atoms in turn reduces the stability of this cluster compared to its neighbors in $x$, and also leaves the cluster ferromagnetic. Similarly, for $n$=4 cluster, Mn$_2$Co$_2$ has larger average Mn-Mn distance (2.83 \AA)  than Mn$_3$Co (2.74 \AA) and found to be less stable. 

\subsection{Magnetic moment}

\begin{figure}[!b]
\includegraphics[width=8.3cm, keepaspectratio, angle=0]{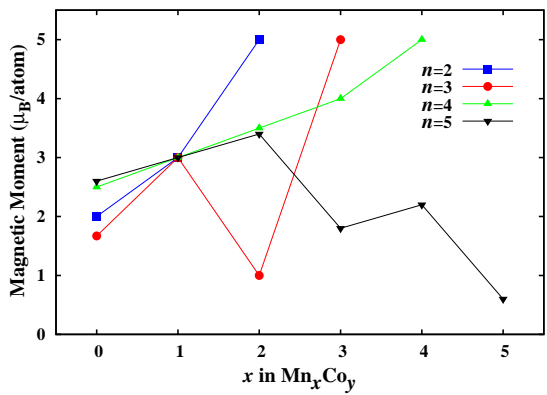}
\caption{\label{fig:mag}(color online) Calculated magnetic moment per atom as a function of Mn-atoms.}
\end{figure}

The atoms in the pure Co$_n$ clusters are found to be aligned ferromagnetically,\cite{Datta2007} whereas a ferromagnetic to ferrimagnetic transition has been observed for pure Mn$_n$ at $n=5$ and remains the same for larger sized clusters.\cite{Kabir2006}  The local magnetic moment of Mn ($\mu_{\rm Mn}$) is found to be higher than $\mu_{\rm Co}$ due to more number of unpaired $d$-electrons in Mn-atom (3$d^5$4$s^2$) than in Co-atom (3$d^7$4$s^2$). 

\begin{figure}[!t]
\includegraphics[width=8.5cm, keepaspectratio, angle=0]{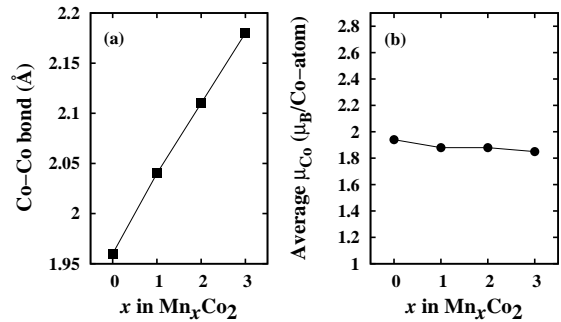}
\caption{\label{fig:Co2mag} (a) Co-Co bond length and (b) average local Co-moment ($\mu_{\rm Co}$) in Mn$_x$Co$_2$ clusters as a function of $x$.}
\end{figure}

Fig.\ref{fig:mag} shows the average magnetic moment $\bar{\mu}$ as a function of the number of Mn-atoms, $x$, for different cluster sizes $n$. For $n$ = 2 and 4,  $\bar{\mu}$ increases with increasing number of Mn-atoms. This is expected since for these sizes, all the clusters with different compositions are ferromagnetic. As mentioned  earlier,   the net number of unpaired $d$-electrons in Mn (five) is greater than that in Co (three). Thus as Co-atoms are replaced by Mn-atoms, total cluster moment increases by $\sim$ 2 $\mu_B$/Mn-substitution. For cluster size $n$=3  and composition $x$=2,  and also for $n$=5  and  $x \ge 3$, i.e., as the clusters become Mn-rich,  $\bar{\mu}$ decreases with $x$.  We see that  these Mn-rich clusters are  ferrimagnetic.  For pure Mn$_n$ clusters ferrimagnetic coupling is observed for $n \ge 5$ (Ref.\cite{Kabir2006}). On the other hand, for  bimetallic Mn$_x$Co$_y$ clusters,  ferrimagnetic coupling of Mn-atoms is observed for $n$=5 with $x \ge 3$.  This could be because, for pure Mn$_{3}$ and Mn$_{4}$, the ferrimagnetic isomers having average magnetic moment 1.67 and 2.50 $\mu_B$/atom, respectively, lie just 0.05 eV and 0.08 eV  above their corresponding ferromagnetic ground states. Thus perturbing these clusters with Co-atoms  could induce the corresponding ground state to be ferrimagnetic. 

\begin{figure*}[!t]
\begin{tabular}{cc}
\begin{minipage}{0.65\textwidth}
\includegraphics[width=9cm, keepaspectratio]{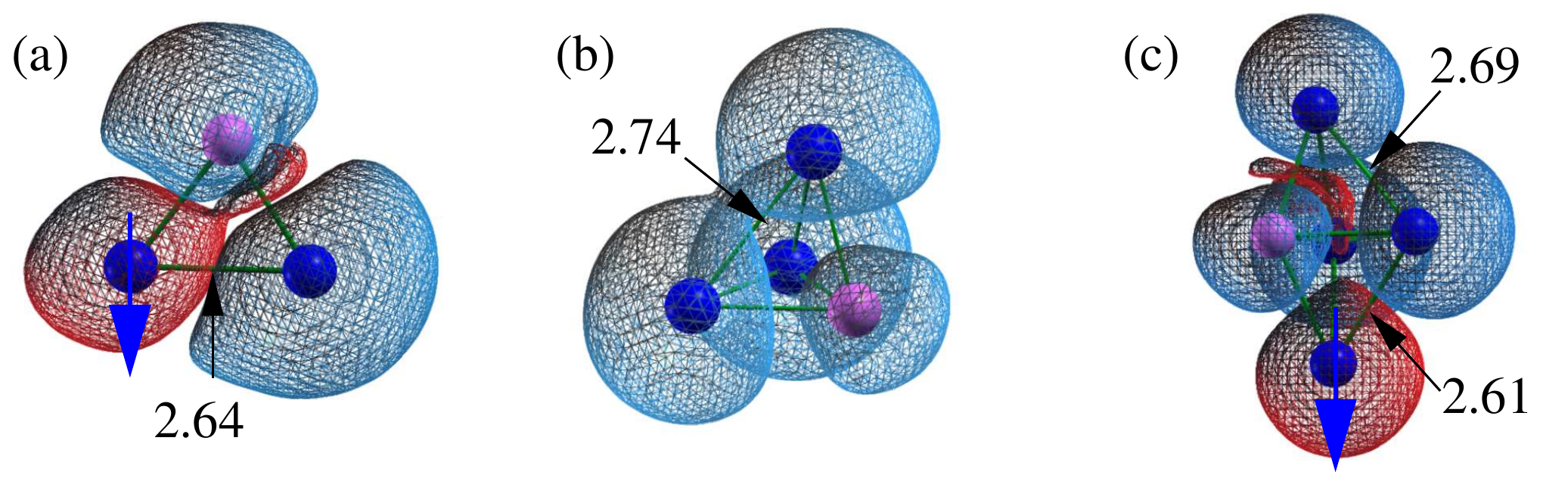}
\end{minipage} &
\begin{minipage}{0.30\textwidth}
\caption{\label{fig:magdens}(Color online) Isosurfaces of magnetization density for (a) Mn$_2$Co, (b) Mn$_3$Co, and (c) Mn$_4$Co corresponding  to 0.02, 0.03, and 0.04 $e$/\AA$^3$, respectively. Dark (Blue) and light  (magenta) color represent Mn- and Co-atoms, respectively. The Mn$_{\downarrow}$ is marked by down arrow. Blue (Red) surface indicates positive (negative)  magnetization density.}
\end{minipage}
\end{tabular}
\end{figure*}

Another interesting observation is made that if we dope  a Co$_2$ dimer with  increasing number of Mn-atoms,  the Co-Co bond length k.pdf increasing (Fig.\ref{fig:Co2mag}a). This indicates that Mn-atoms reduce the  hybridization between Co-atoms in the dimer.  Therefore, we have further studied the average $\mu_{\rm Co}$ in Co$_2$ dimer with increasing $x$,  and is shown in Fig.\ref{fig:Co2mag}b. It is seen that $\mu_{\rm Co}$ remain almost the same for all the Mn$_x$Co$_2$ clusters (1.94, 1.88, 1.88,  and 1.85 $\mu_{B}$/atom for Co$_{2}$, MnCo$_{2}$,  Mn$_{2}$Co$_{2}$,  and Mn$_{3}$Co$_{2}$, respectively).  Although the Co-Co bond length in Mn$_x$Co$_2$ increases with increasing $x$,  the coordination of Co-atoms also increases as we go   along Co$_{2}$ $\rightarrow$  MnCo$_{2}$ $\rightarrow$ Mn$_{2}$Co$_{2}$ $\rightarrow$ Mn$_{3}$Co$_{2}$,  which leaves the effective hybridization of the Co-atoms unaffected.

The magnetization density (difference between the up and down spin densities)  further illustrates (Fig.\ref{fig:magdens}) the magnetic nature of these  nanoalloy clusters. It is interesting to note that in ferrimagnetic Mn$_2$Co (Mn$_4$Co) cluster the Mn$_{\uparrow}$-Mn$_{\downarrow}$ separation is 2.64 \AA \ (2.61 and 2.65 \AA). This is significantly smaller than Mn$_{\uparrow}$-Mn$_{\uparrow}$ separation in  ferromagnetic Mn$_3$ and Mn$_4$ clusters, 2.74 and 2.70, respectively.\cite{Kabir2006} Also the average Mn$_{\uparrow}$-Mn$_{\uparrow}$ bond in ferrimagnetic Mn$_4$Co  is longer (Fig.\ref{fig:magdens}c) than the Mn$_{\uparrow}$-Mn$_{\downarrow}$ one. Thus it seems that wherever there is a contraction of Mn-Mn bond   due to Co doping,  the two Mn atoms concerned get antiferromagnetically  aligned.  This is further supported by the fact  that for Mn$_3$Co, where the distances between the Mn  atoms (2.74 \AA) is not reduced by Co-doping, the Mn atoms  remain ferromagnetically coupled.  The dependence of Mn-Mn coupling on the separation is due to the  modification of hybridization. We will further discuss the magnetic nature of the clusters in  Section \ref{sec:pcd} through partial charge density.

\subsection{Comparison of magnetic moment with Stern-Gerlach experiment}

\begin{figure}[!b]
\includegraphics[width=8.3cm, keepaspectratio, angle=0]{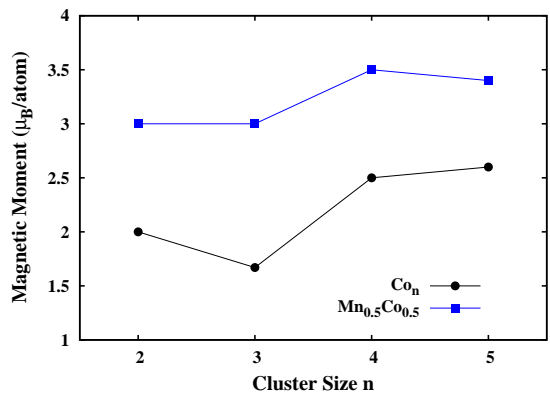}
\caption{\label{fig:mag_comp}(color online) Magnetic moment of pure Co$_n$ and Mn$_{0.5}$Co$_{0.5}$ clusters. For cluster sizes $n=3$ and 5, the cobalt rich clusters (MnCo$_2$ and Mn$_2$Co$_3$, respectively) are considered. See the text.}
\end{figure}

In a recent SG experiment,  Mn$_x$Co$_y$ ($n$=11-29) clusters were  produced via pulsed laser vaporization from cylindrical Mn$_x$Co$_{1-x}$  alloy (Mn$_{0.15}$Co$_{0.85}$ or Mn$_{0.5}$Co$_{0.5}$) target rods.\cite{Mark2007}   The {\it mass spectrum} shows that average Mn-fraction in the alloy  clusters are always less than in the corresponding source rod. However, this  Mn-fraction increases with the increase in cluster size. Therefore, in the experiment, alloy clusters were always Co-rich compared to the bulk source.\cite{Mark2007} Clusters generated from both   the Mn$_{0.15}$Co$_{0.85}$ and Mn$_{0.5}$Co$_{0.5}$ target rods have moments  that are similar to that of  the pure cobalt clusters of same size. In fact in {\it smaller} size range, $n$=11-14,  the average moment is enhanced by 28-60\% for clusters generated from  Mn$_{0.50}$Co$_{0.50}$ source. We have compared the magnetic moments of pure Co$_n$ clusters with 50-50 Mn-Co nanoalloy clusters (Fig.\ref{fig:mag_comp})  in the size range $n=$2-5 to see whether  the observed experimental trend continues for the smaller sizes.  It should be pointed out here that, for Mn$_{0.50}$Co$_{0.50}$ clusters, we have considered the Co-rich clusters to mimic the experimental  situation,\cite{Mark2007} for sizes where an exact 50-50 concentration is not possible.   We see a 80-31\% enhancement in magnetic moment compared to pure Co$_n$ clusters.  As we have already seen that these nanoalloy clusters are ferromagnetic  unless the cluster is Mn-rich, thus when we replace Co-atoms in pure  Co$_n$ clusters with Mn, the magnetic moment increases. 

In a separate SG experiment,\cite{Yin2007} Yin {\it et al.} has found that magnetic moment  increases as the number of Co-atoms, $y$, in Mn$_x$Co$_y$ increases  in the range 15-45 keeping $x$ constant at $1-9$ (i.e., all the cluster are Co-rich).  This observation also holds good in the present study for smaller clusters as long as the alloy cluster is Co-rich. For an example,  as we go   along MnCo (6 $\mu_B$) $\rightarrow$ MnCo$_2$ (9 $\mu_{B}$) $\rightarrow$  MnCo$_3$ (12 $\mu_{B}$) $\rightarrow$ MnCo$_4$ (15 $\mu_{B}$)  the total moment increases with increasing $y$. They have also found that the total enhancement in the magnetic moment  to be 1.7 $\mu_B$/Mn-substitution, and is independent of the cluster size  $n$ and composition, which is justified by the virtual bond state  model.\cite{Yin2007} The present calculation is in good agreement and we find this enhancement to be 2  $\mu_{B}$/Mn-substitution (Tables \ref{table:n=2}-\ref{table:n=5}) as long as cluster is ferromagnetic (Co-rich).

\subsection{Comparison with bulk alloy}

The individual atomic moments of  Mn and Co for Mn$_{x}$Co$_{1-x}$ bulk alloy have been determined with increasing Mn:Co ratio through neutron scattering studies\cite{Mark2007}. On the Co-rich side $\mu_{\rm Mn}$ and  $\mu_{\rm Co}$ are aligned antiferromagnetically,  with the  magnitude of both $\mu_{\rm Mn}$ and $\mu_{\rm Co}$ decreasing  monotonically with increasing Mn-content. Same trend is  also seen for the  average magnetic moment of the alloy $\bar{\mu}$=$x \mu_{\rm Mn}$ +$(1-x) \mu _{\rm Co}$.   However, it is seen from the SG experiments\cite{Mark2007,Yin2007} that Co-rich Mn$_x$Co$_{1-x}$ nanoalloy clusters retain substantial magnetic moment in the size range $n$=11-29.  Moreover, for clusters with $n$=11-14 obtained by laser vaporization of  Mn$_{0.15}$Co$_{0.85}$ rods, the magnetic moment increases by 88-148\%  compared to the bulk value.\cite{Mark2007} At 20\% Mn-concentration for cluster size $n$=5 we estimate  241\% enhancement in magnetic moment from the corresponding bulk value. This enhancement of $\bar{\mu}$ in low dimension over bulk alloy  is not only due to decrease in coordination number, which effectively  reduces the hybridization among the orbitals, but also due to the  ferromagnetic Mn-Co coupling (unlike bulk alloy)  in Co-rich Mn$_x$Co$_y$ nanoalloy clusters.  

\begin{figure}[!t]
\includegraphics[width=8cm, keepaspectratio, angle=0]{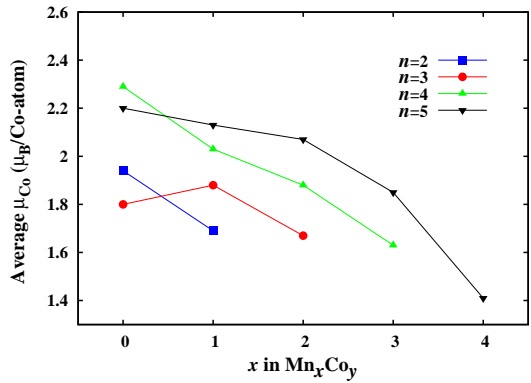}
\caption{\label{fig:locComag}(color online) Average local Co-moment ($\mu_{\rm Co}$) as a function of Mn-atoms for different cluster sizes.}
\end{figure}

The average  $\mu_{\rm Mn}$ and $\mu_{\rm Co}$ (calculated using Eq. 2) are studied individually for different sizes with increasing Mn-concentration. The average $\mu_{\rm Mn}$ oscillates with increasing $x$. For clusters like Mn$_2$Co which are ferrimagnetic, the average magnetic moment of Mn is very low (0.08$\mu_{B}$/atom),  whereas for ferromagnetic (like Mn$_2$Co$_2$) clusters it is high (4.21$\mu_{B}$/atom). Here, it is to be noted that even for such ferrimagnetic clusters the individual $\mu_{\rm Mn}$ values are high but due do their  ferrimagnetic Mn-Mn alignment the net Mn-moment is low. The average $\mu_{\rm Co}$ is plotted in Fig.\ref{fig:locComag}. It is clear from Fig.\ref{fig:locComag} that, likewise in bulk alloy,  the average $\mu_{\rm Co}$ decreases with increasing Mn concentration for all cluster sizes,\cite{Co3}  and, as we will see in Section \ref{sec:pdos}, is because of increasing Mn neighbor. However, $\mu_{\rm Mn}$ doesn't behave in similar fashion, and unlike the bulk alloy, in Co-rich Mn$_x$Co$_y$  nanoalloy clusters, $\mu_{\rm Mn}$ and $\mu_{\rm Co}$ are ferromagnetically coupled.

\subsection{\label{sec:n13}A medium sized cluster ($n=$13) and direct comparison with SG experiment}

\begin{figure}[t]
\includegraphics[width=8cm,  keepaspectratio]{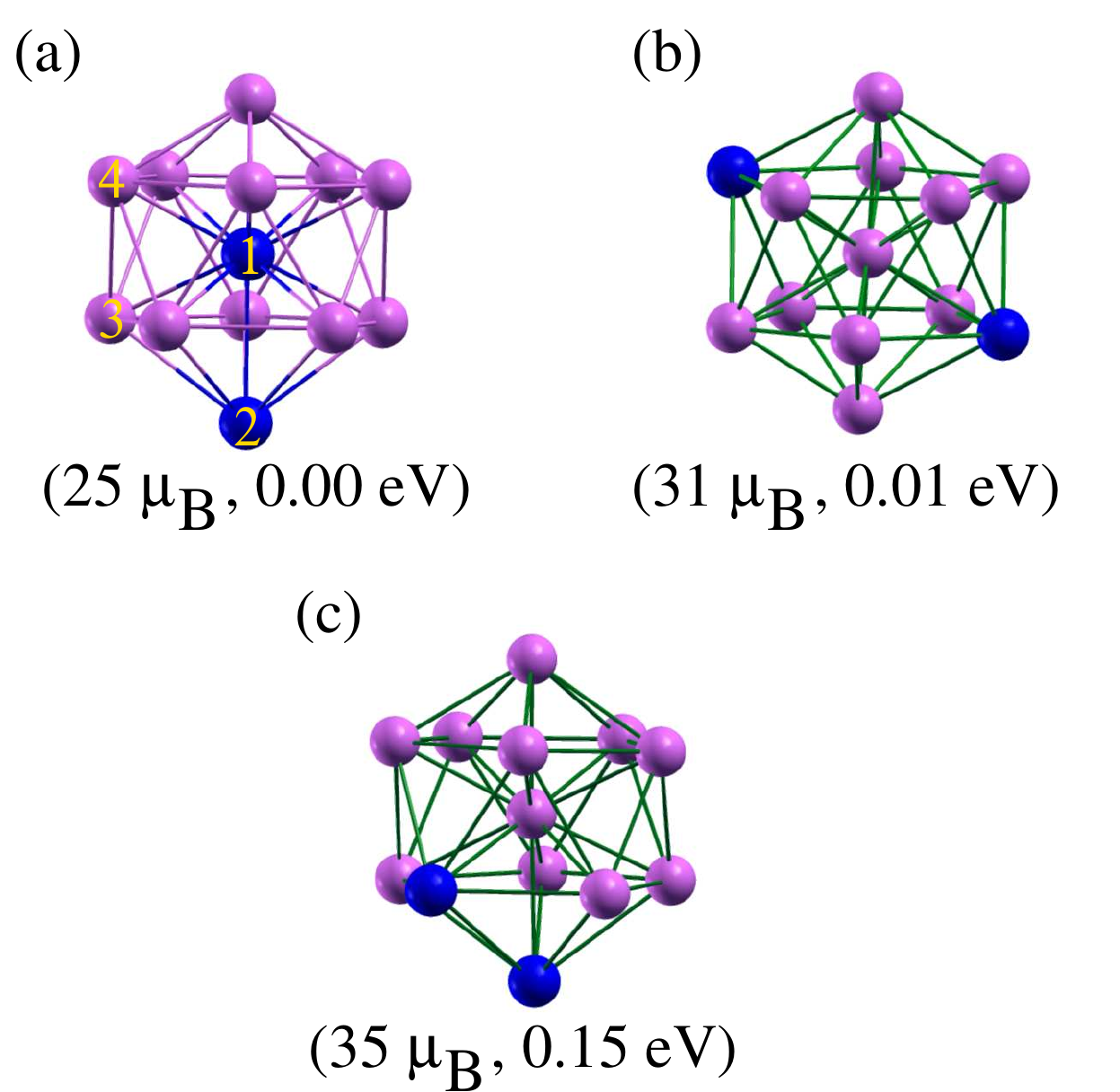}
\caption{\label{fig:Mn2Co11}(Color online) The ground state (a),  first (b), and second (c) isomers for Co$_{11}$Mn$_{2}$ nanoalloy cluster. Same color conventions are used as in Fig.\ref{fig:trimer}. The numbers in the parenthesis represent the total moment and relative energy to the ground state, $\Delta E$, respectively.}
\end{figure}

In order to perform a direct comparison with the SG experiment\cite{Mark2007} we have studied a  Mn$_x$Co$_y$ cluster of size $n$=13, with 15\% Mn-concentration i.e.,  Mn$_2$Co$_{11}$ nanoalloy cluster.  The ground state is found to be nearly  degenerate with magnetic moments  1.92 and 2.38 $\mu_B$/atom. The ground state has binding energy 3.18 eV/atom  (Fig.\ref{fig:Mn2Co11}a), whereas the nearly degenerate isomer lies only 0.01 eV  higher (Fig.\ref{fig:Mn2Co11}b).    In the first structure the Mn-Mn distance is 2.46 \AA \ and one of the Mn-atom sits at the center, and consequently the magnetic moment of this  (Mn) atom gets quenched.   Whereas, the Mn-Mn distance is comparatively  larger (4.78 \AA) in the second cluster as both the Mn-atoms sit on the surface. These are the  reasons that  the second structure has larger magnetic moment than the first one.   Another icosahedral isomer (Fig.\ref{fig:Mn2Co11}c) has large magnetic moment of 2.69 $\mu_B$/atom and lies 0.15 eV higher than the ground state.

The icosahedral ground state structure for this Mn-doped cluster is quite interesting. For pure Co$_{13}$ the stable ground state is a distorted  hexagonal structure\cite{Datta2007} and, on the other hand,  we have previously found an icosahedral ground state for pure Mn$_{13}$  cluster.\cite{Kabir2006,Kabir2007} Replacing two Co-atoms in Co$_{13}$ with Mn results in a structural change from a distorted-hexagonal-like structure to an icosahedron. The calculated magnetic moment of the pure Co$_{13}$ is found to be 25 $\mu_B$,  which is in good agreement with experimental values, 26-30 $\mu_B$ (Refs. \cite{Mark2006, Xu2005}).  On the other hand, two degenerate ground states of Mn$_2$Co$_{11}$ nanoalloy cluster  have 25-31 $\mu_B$ magnetic moments, which is 0-6 $\mu_B$ larger than that of (calculated) the pure Co$_{13}$ cluster. This is in good agreement with the experiment: For   $n$=13 cluster produced from Mn$_{0.15}$Co$_{0.85}$ bulk source an enhancement  in average magnetic moment  is seen over the corresponding  pure Co$_{13}$ cluster.\cite{Mark2006} It should also be pointed out here that the SG experiment was done at finite temperature (91 K), so the cluster beam may contain both the degenerate states which are separated only by 0.01 eV energy.

\subsection{\label{sec:pdos}Projected density of states}

\begin{figure}[!t]
\includegraphics[width=9.5cm,  keepaspectratio, angle=0]{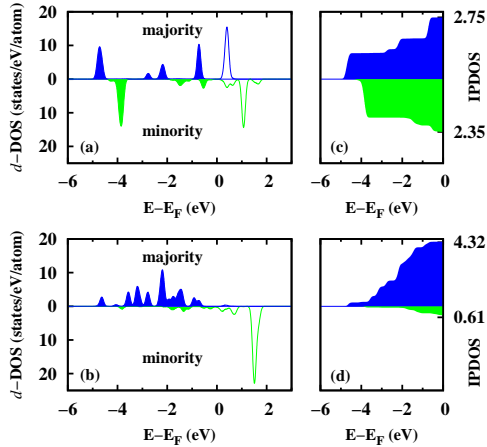}
\caption{\label{fig:dos}(Color online) The $d$-projected density of states for the (a) central, and (b) surface Mn-atom of Mn$_2$Co$_{11}$ cluster in its ground state (marked as 1, and 2, respectively, in Fig.\ref{fig:Mn2Co11}a). A  Gaussian broadening of 0.1 eV has been used. Corresponding integrated density of states up to the Fermi level, $E_F$, are also shown for the (c) central, and (d) surface Mn-atoms.}
\end{figure}

To understand the decrease in average $\mu_{\rm Co}$  when the environment is made more Mn-rich than Co (Fig.\ref{fig:locComag}), the  integrated projected density of states (IPDOS) of the $d$-orbital has been investigated for the cobalt atoms in Mn$_2$Co$_3$ and Mn$_4$Co clusters.  For the first cluster in the TBP ground state (Fig.\ref{fig:pentamer}),  the IPDOS of the Co-atom, which is co-planer with two Mn-atoms,  is 4.41  for the  majority spin channel, whereas it is 2.30 for the minority spin.  For a off-plane Co-atom in same structure, the IPDOS  is 4.30 for the majority spin channel and 2.40 for the minority spin one.  The reason for this greater spin polarization of the co-planar Co-atom  is twofold. Firstly, the in-plane Co-atom has  more (two) Co-atoms as its nearest neighbors than the  off-planar ones (one).  Secondly, the in-plane Mn-Co distance  (2.62 \AA) is larger than that for the off-planar Mn-Co distance (2.41 \AA). Thus the effect of Mn-atoms felt by the  in-plane Co-atom is less than that by the off-planar ones.  This reinforces the observation  that a Co-rich environment favors more spin polarization in Co than  a Mn-rich environment.  For the Co-atom in the second cluster, Mn$_4$Co, the  IPDOS for the majority (minority) spin channel is 3.81 (2.50),  giving rise to a decrease  in spin polarization (compared to average $d$-polarization of Co-atoms  in Mn$_2$Co$_3$), when the environment is  made more Mn-rich at the cost of Co.      

The $d$-projected density of states ($d$-DOS) and the corresponding  IPDOS for Mn$_2$Co$_{11}$ are also studied. These are shown in Fig.\ref{fig:dos} for the central, and apex Mn-atoms labeled  as 1 and 2, respectively,  in  Fig.\ref{fig:Mn2Co11}a. We see  that for the central Mn-atom (Fig.\ref{fig:dos}a), the  spin distribution in the majority and minority spin  channels are nearly equal. Consequently, the magnitude of  IPDOS   (Fig.\ref{fig:dos}c) for the majority and minority spin channels  are close, 2.75 and 2.35, respectively.   Thus for this central Mn-atom spin   polarization is very low.  On the other hand  for the surface Mn-atom (Fig.\ref{fig:dos}d), the majority spin channel is almost  fully occupied (Fig.\ref{fig:dos}b)  and the magnitude of corresponding IPDOS for the majority and minority spin  channels are 4.32 and 0.61, respectively.  This gives a large spin polarization on this atom.  This is because the surface Mn is less coordinated and hence has less hybridization than the central one. Further, the Co-atoms marked as 3 and 4 in Fig.\ref{fig:Mn2Co11}a have equal coordination number and the former has slightly larger (2.43 \AA) average bond length than  the later one (2.38 \AA). Although we find that the Co-atom marked as 3  has slightly weaker spin polarization (1.74 $\mu_B$) than that of marked as 4 (1.83 $\mu_B$), as the former is bonded with one more Mn-atom than the later. This once again confirms our observation that $\mu_{\rm Co}$ decreases  with increasing Mn neighbors.

\begin{figure}[!t]
\includegraphics[width=8.4cm,  keepaspectratio]{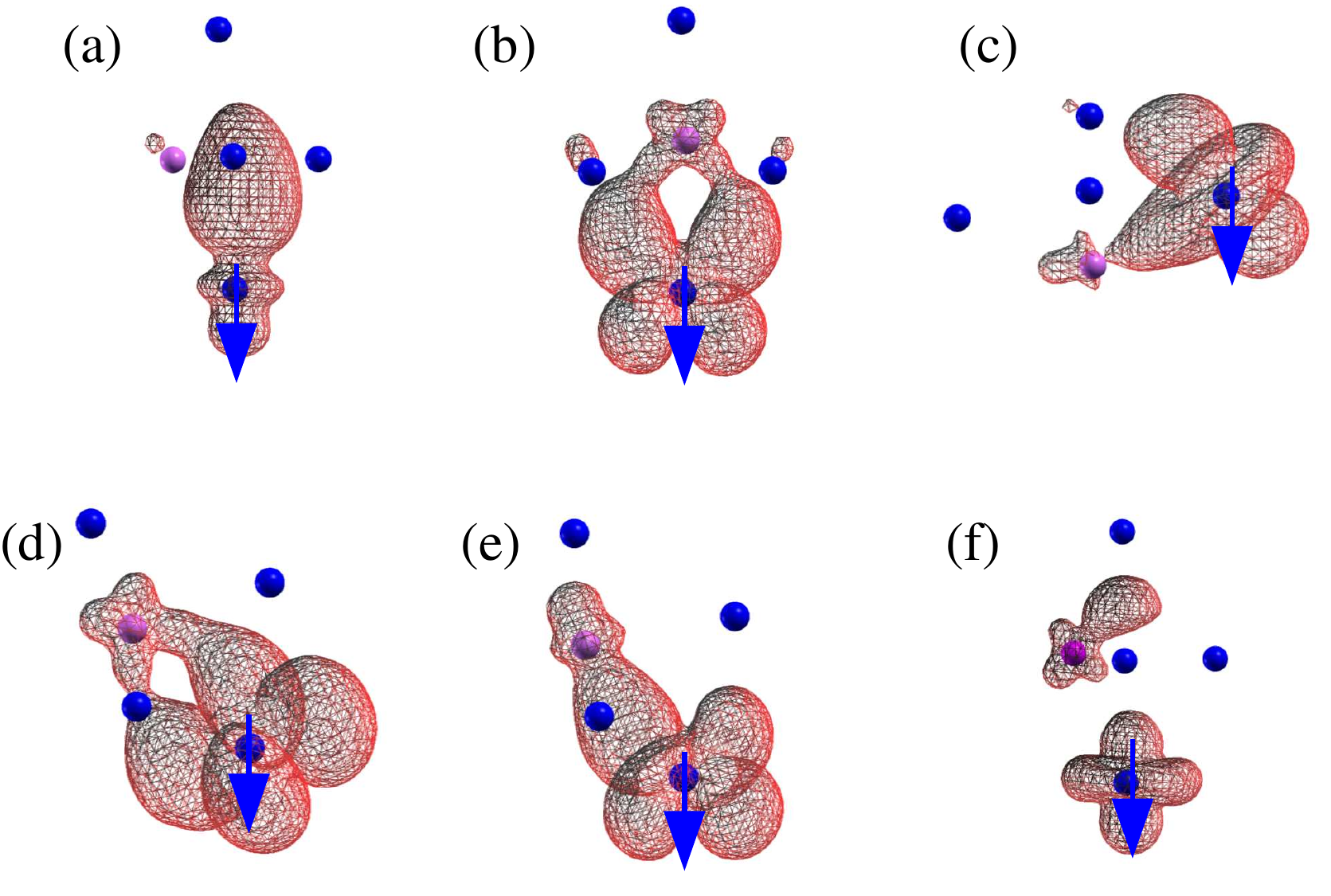}
\caption{\label{fig:chgdenMn4Co-minority}(Color online) The charge density  isosurfaces of the six down electrons in the deepest energy levels for Mn$_4$Co cluster. They are shown at (a) 0.04, (b) 0.01, (c) 0.01, (d) 0.01, (e) 0.01,  and (f) 0.03 $e$/\AA$^3$, respectively.  Dark (Blue) and light  (magenta) color represent Mn- and Co-atoms, respectively,  and the  Mn$_{\downarrow}$ is marked by down arrow.}
\end{figure}

\subsection{\label{sec:pcd}Partial charge densities}

\begin{figure}[!t]
\includegraphics[width=7cm,  keepaspectratio]{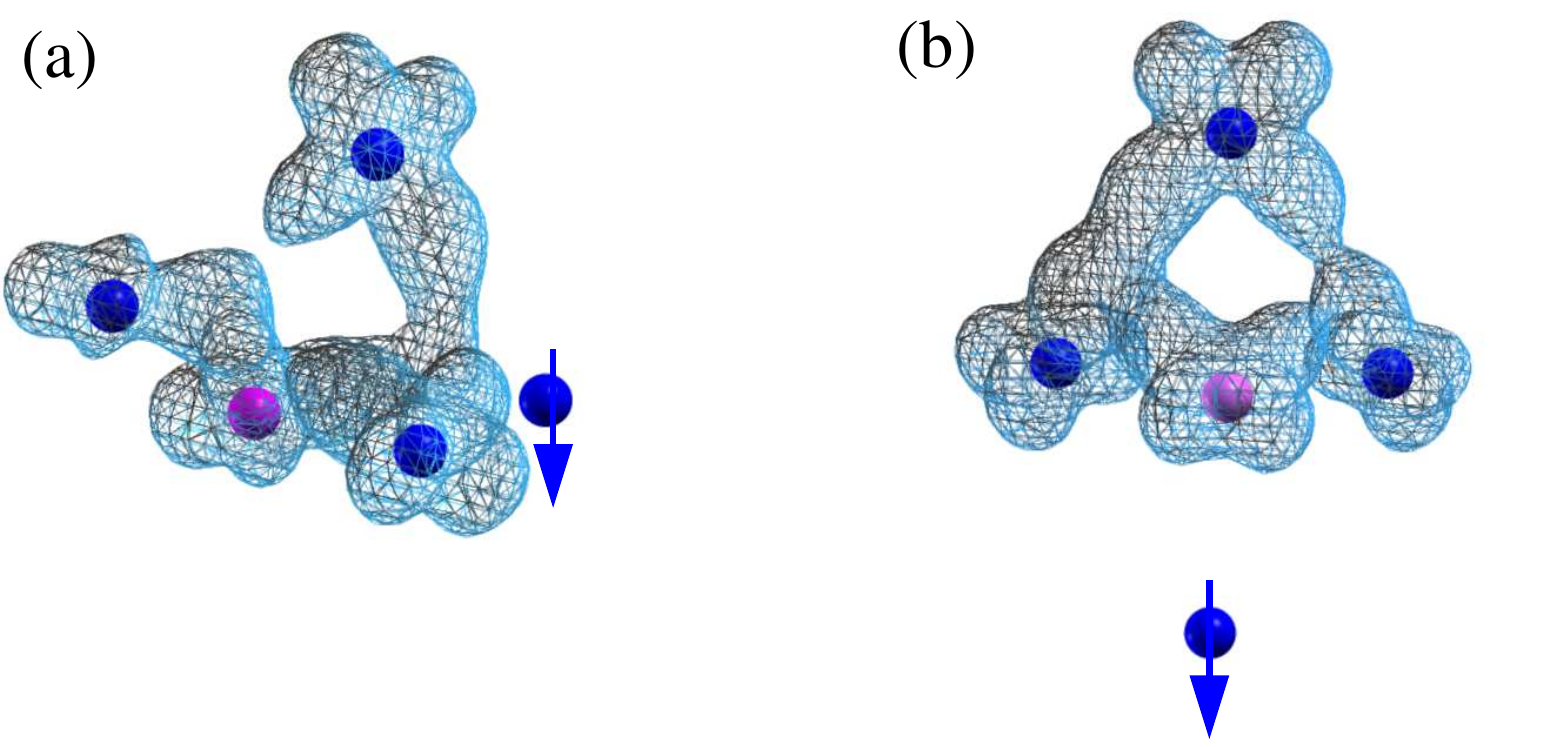}
\caption{\label{fig:chgdenMn4Co-majority}(Color online) The charge density distribution for two of the majority electrons in the deep levels for Mn$_4$Co cluster.  Dark (Blue) and light (magenta) color represent Mn- and Co-atoms,  respectively, and the Mn$_{\downarrow}$ is marked by down arrow. The  isosurfaces are drawn at 0.04  $e$/\AA$^3$ density.}
\end{figure}

 \begin{figure}[!t]
\includegraphics[width=7cm,  keepaspectratio]{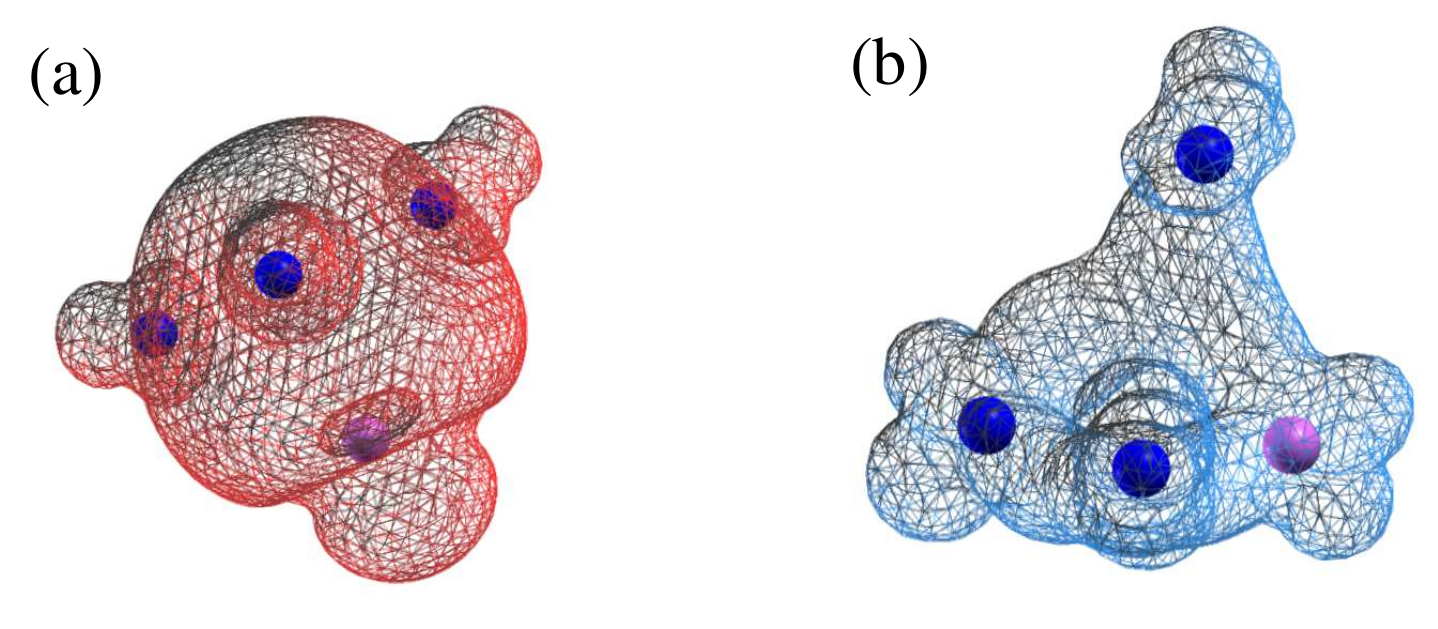}
\caption{\label{fig:chgdenMn3Co} (Color online) The charge density  distribution for (a) the minority electron at 0.01 e/\AA$^{3}$ isodensity  and (b) for one of the majority electrons at  0.03 e/\AA$^{3}$ isodensity  in the deep levels of Mn$_{3}$Co cluster. Dark (Blue) and light  (magenta) color represent Mn- and Co-atoms, respectively.}
\end{figure}

In order to take a deeper look into the electronic behavior responsible for  antiparallel alignment of  one $\mu_{\rm Mn}$ to other Co- and Mn-atoms  in Mn$_4$Co, we have investigated the eigenvalue spectrum of the occupied levels.  We find among the eight highest occupied levels (i.e., HOMO $ - $ $n'$ levels, where $n'=0-7$), four of them are occupied by minority spins. On the other hand, among the eleven deepest occupied levels, six levels are occupied by minority  electrons. Comparing this spin distribution picture with that of Mn$_3$Co, we see that for this ferromagnetic cluster also,  only four of the eight highest occupied levels are occupied by minority spins likewise in the ferrimagnetic  Mn$_4$Co. However, only one of the deepest eleven states are occupied with  minority spin. Therefore, it seems that these deep valence levels play a major  role in determining the magnetic structure of these two clusters.

We have further looked at  the charge density distribution of the  six minority spin electrons in ferrimagnetic Mn$_4$Co nanocluster, which occupy these  deepest molecular orbitals. At large isosurface values the  charge densities are localized on the Mn$_{\downarrow}$ and are  distributed mostly between this Mn$_{\downarrow}$ and the Co-atom
for lower isosurface values (Fig.\ref{fig:chgdenMn4Co-minority}).     The shape of the charge densities indicates that these electrons  are $d$-electrons.  Thus, single Co-doping in ferromagnetic Mn$_4$  makes Mn$_4$Co ferrimagnetic.   On the other hand, the charge densities of the majority electrons  are spread out among the Co-atom and the three Mn$_{\uparrow}$-atoms  at small isosurface values. Charge densities for two of these majority  electrons are shown in Fig.\ref{fig:chgdenMn4Co-majority}.     The shape of these charge density isosurfaces  also exhibits $d$-orbital character. It is interesting to observe that the  Mn$_{\downarrow}$ is completely devoid of majority charge contribution for these 
deep valence levels.  Thus it must be so that the  Mn$_{\downarrow}$ goes from the 4$s^{2}$3$d^{5}$ to 4$s^{1}$3$d^{6}$ configuration in Mn$_{4}$Co. Due to Co doping the Mn$_{\downarrow}$-Mn$_{\uparrow}$ separation in the cluster is  reduced, which consequently enhances  the hybridization  and cause the six $d$-electrons in  this Mn$_{\downarrow}$ to belong to the minority spin channel.

In contrast the charge density of the electrons occupying the deepest energy levels of Mn$_3$Co cluster exhibits quite a different picture. The charge densities of both the majority and minority electrons for small isosurface value (Fig.\ref{fig:chgdenMn3Co}) are spread  out amongst all the four atoms in the cluster and not localized on  any one of them.

\section{Summary}
 
We have studied structure, bonding, and magnetism in small bimetallic Mn$_x$Co$_y$ ($x+y$=2-5) clusters from first-principles DFT calculation. Due to van der Waals {\it like} weaker bonding among Mn-atoms and relatively strong Co-Co bonding than Mn-Co bonding, the binding energies of the alloy clusters decrease with increasing Mn-concentration.  Interesting effects in binding energy, stability and  magnetism in the nanoalloy  clusters are explained through the interplay between bond length and coordination. The Co-rich clusters are found to be ferromagnetic unlike the bulk  alloy and the corresponding magnetic moment is higher than the pure Co$_n$ clusters as is seen in the recent SG experiments.\cite{Mark2007, Yin2007}  Moreover, the magnetic moment of Co-rich nanoalloy clusters increase with Mn-concentration and this  increment is  2 $\mu_B$/Mn-substitution and is independent of cluster size and composition.   Co-atoms are found to be more magnetically polarized in  a Co-rich environment than in Mn-rich one, i.e., likewise in bulk alloy, as the environment  is made more Mn-rich, the average $\mu_{\rm Co}$ decreases.

\acknowledgements
We thank D. G. Kanhere for stimulating discussion. M. K. thanks M. B. Knickelbein for sharing his experimental data before it was published as Ref.\cite{Mark2007}, which greatly motivated this work. S. D. thanks CSIR for financial support. B. S. and A. M. acknowledge Asian-Swedish Research Links Programme for financial support.


\begin{thebibliography}{99}
\bibitem{stable}A spin arrangement in any magnetic clusters is magnetically stable 
only if the lowest unoccupied molecular orbital (LUMO) of the majority spin lies 
above the highest occupied molecular orbital (HOMO) of the minority spin and vice
versa.
\bibitem{Co3}It should be noted that the pure Co$_3$ has two degenerate ground
states with 1.67 and 2.33 $\mu_B$/atom.
\end{thebibliography}
\end{document}